\documentstyle[aps,prb,floats,twocolumn]{revtex}
\input epsf     

\begin{document}
\draft
\title{Invertible and non-invertible alloy Ising problems}
\author{C. Wolverton and Alex Zunger}
\address{National Renewable Energy Laboratory, Golden, CO 80401\\}
\author{B. Sch\"onfeld}
\address{Institut f\"ur Angewandte Physik, 
Eidgen\"ossische Technische Hochschule Z\"urich,
CH-8093 Z\"urich, Switzerland\\}
\date{\today}
\maketitle
{\let\clearpage\relax
\twocolumn[%
\widetext\leftskip=0.10753\textwidth \rightskip\leftskip
\begin{abstract}
Physical properties of alloys are compared as computed
from ``direct'' and ``inverse'' procedures.  The direct
procedure involves Monte Carlo simulations of a set of 
local density approximation (LDA)-derived pair
and multibody interactions $\{ \nu_f \}$, generating short-range
order (SRO), ground states, order-disorder transition temperatures,
and structural energy differences.
The inverse procedure involves
``inverting'' the SRO generated from $\{ \nu_f \}$
via inverse-Monte-Carlo
to obtain a set of {\em pair only}
interactions $\{ \tilde{\nu}_f \}$.
The physical properties generated from
$\{ \tilde{\nu}_f \}$
are then compared with those from $\{ \nu_f \}$.
We find that (i) inversion of the SRO is possible 
(even when $\{ \nu_f \}$ contains multibody interactions
but $\{ \tilde{\nu}_f \}$ does not) but,
(ii) the resulting interactions $\{\tilde{\nu}_f\}$
agree with the input interactions
$\{\nu_f\}$ only when the problem is dominated by
{\em pair} interactions.  Otherwise, $\{\tilde{\nu}_f\}$ are
{\em very} different from $\{\nu_f\}$.
(iii) The same SRO pattern
can be produced by drastically different sets
$\{\nu_f\}$.  Thus, the effective interactions deduced from
inverting SRO are not unique.
(iv) Inverting SRO always misses configuration-independent (but
composition-dependent) energies such as the 
volume deformation energy $G(x)$; consequently,
the ensuing $\{\tilde{\nu}_f\}$
cannot be used to describe formation enthalpies or
two-phase regions of the phase diagram, which depend
on $G(x)$.
\end{abstract}
\vspace{11pt}

]}
\narrowtext

The physical properties of $A_{1-x}B_x$ alloys are usually analyzed
and interpreted
via ``cluster expansion'' models: \cite{Ducastelle91,Zunger94,deFontaine94}
Each of the $N$ sites of an alloy $i$=1,...,$N$ is labeled by
a spin variable $\hat{S}_i$ = -1 or +1 if site $i$ is occupied
by an $A$ or $B$ atom, respectively.
The set of spin variables $\{\hat{S}_i\}$ defines a
configuration $\sigma$.  The energy of any of the $2^N$ possible
configurations is then written as a sum over clusters of 
points $\{i;ij;ijk;...\}$:  \cite{SDG}
\begin{eqnarray}
\label{ising}
E(\sigma,V)  =  J_0(V) &+& \sum_{i} J_i(V) \hat{S}_i
+ \sum_{j<i} J_{ij}(V) \hat{S}_i\hat{S}_j \nonumber \\
&+& \sum_{k<j<i} J_{ijk}(V) \hat{S}_i\hat{S}_j
\hat{S}_k + ...
\end{eqnarray}
where $V$ is the volume, the $J$'s are interaction energies and
the first sum is over all sites, the second over all pairs, the
third over all triplets, etc.  We refer to these elementary
clusters as ``figures'' $f$.  

If the set of interactions $\{ J_f(V) \}$ is
known for a given alloy system, 
one may apply
standard methods of lattice statistical mechanics
(e.g., mean field, cluster variation, or Monte Carlo methods)
to the expansion
and compute ground state structures or finite-temperature
thermodynamic properties.
Recent examples include the calculation of
temperature-composition phase diagrams and ground state
structures of transition metal
\cite{Ducastelle91,Zunger94,deFontaine94} 
and semiconductor \cite{Zunger94} alloys, mixing enthalpies of
disordered, partially ordered, and off-stoichiometric alloys, 
\cite{Zunger94,deFontaine94} 
and short-range order (SRO) of solid solutions.
\cite{Lu94,Wolverton95} 
We refer to this approach as the {\em ``direct approach''}.

Conversely, another common tradition involves the 
{\em ``inverse approach''}:
A measured thermodynamic property such as the set of SRO
parameters ${\alpha(n)}$
(the atom-atom pair correlation for the $n$th atomic shell)
is used in an inverse statistical approach
(e.g., the inverse Monte Carlo (IMC) method \cite{Gerold87})
to deduce a set of effective interactions.  \cite{Schweika88}
These interactions are subsequently used in a cluster
expansion [Eq. (\ref{ising})] to predict thermodynamic
properties other than the SRO.
In this paper, we explore the extent to which
the inverse approach may be used to predict alloy properties
by applying it to a well-characterized $\alpha(n)$ obtained
through a direct procedure.

%
%
\begin{table*}
\caption{
The values of the input interaction energies $\nu_f$
and the interaction energies $\tilde{\nu}_f$ reconstructed
via IMC simulations of the SRO computed from $\nu_f$
(meV/atom).
Designation of the ``figures'' $f$
follows the notation of Table IV of Ref. \protect\onlinecite{Lu94}.
For Set 2, the multibody interactions used in the direct set
$D_f\nu_f$, but not in the inverse set
$D_f\tilde{\nu}_f$ are (in meV/atom):
$J_3$=$-$96.1, 
$K_3$=44.5, 
$L_3$=64.5, 
$M_3$=$-$41.1,
$Q_3$=$-$81.3, and 
$K_4$=139.1.
Structural energy differences,
ordering energies $\delta E_{\rm ord}(\sigma)$
(the energy difference between
$\sigma$ and a random alloy at the same composition),
and the random alloy mixing energy at $x$=1/4 are
shown (meV/atom), as are
transition temperatures (K).
``NA'' means not applicable.}
\label{comp.jf}
\begin{tabular}{ccdddd}
&&\multicolumn{2}{c}{Set 1} & 
\multicolumn{2}{c}{Set 2}\\

&& Direct & Inverse & Direct & Inverse\\
\tableline

Clusters&Designation & $D_f\nu_f$ & $D_f\tilde{\nu}_f$ 
            & $D_f\nu_f$ & $D_f\tilde{\nu}_f$ \\

Empty & $J_0$ & $-$233.2 &  NA     &$-$233.2 &  NA  \\
Point & $J_1$ &    252.9 &  NA     &   252.9 &  NA  \\
Pairs & $J_2$ &    152.0 & 157.2 &   152.0 & 690.0  \\
&       $K_2$ &  $-$20.0 &$-$21.0& $-$20.0 &  17.6  \\
&       $L_2$ &     58.9 &  60.0 &    58.9 &$-$19.2 \\
&       $M_2$ &     33.5 &  33.3 &    33.5 & 103.2  \\
&       $N_2$ &          &       &     0.0 &$-$3.6  \\
&       $O_2$ &          &       &     0.0 &$-$0.4  \\
&       $P_2$ &          &       &     0.0 &  13.2  \\
\tableline
&$\delta E(L1_2,D0_{22})$      &   -4.0   &  -4.8 & +103.3  &+76.2   \\
&$\delta E_{\rm ord}(L1_2)$    &  -42.6   & -45.1 &  -41.9  &-79.8   \\
&Ground State                  &  $L1_2$  &$L1_2$ &$D0_{22}$&$D0_{22}$\\
&$T_c$                         &  630     &  680  &  1850   &  1900  \\
&$G(x=1/4)$                    &  112.1   &   NA  &  112.1  &   NA   \\
&$\Delta H_{\rm mix}(1/4)$&-56.2  &-172.1 & -227.7  & -600.6 \\
\end{tabular}
\end{table*}

In the following it is convenient to introduce the {\em excess}
energy $\Delta E(\sigma,V)$ of configuration $\sigma$ defined as the
energy of this configuration at volume $V$, relative to
the energies $E_A(V_A)$ and $E_B(V_B)$ of equivalent
amounts of solid $A$ and $B$, at their respective 
equilibrium volumes $V_A$ and $V_B$:
\begin{equation}
\label{deltae}
\Delta E(\sigma,V) = E(\sigma,V) - [ (1-x)E_A(V_A)
+ xE_B(V_B) ].
\end{equation}
If the equilibrium volume $V(\sigma)$ depends
primarily on the composition $x$ and only weakly on the
configuration $\sigma$, then the variables $\sigma$
and $x$ can be rigorously separated in Eq. (\ref{deltae})
giving \cite{Ferreira88}
\begin{equation}
\label{eps-G}
\Delta E(\sigma,V) = G(x) + 
\sum_{f} D_f \hspace{3pt} \nu_f \hspace{3pt}
\overline{\Pi}_f(\sigma).
\end{equation}
Here, the first term $G(x)$ describes the elastic energy 
necessary to deform the constituents from their
equilibrium volume to the volume $V(x)$ of $\sigma$.
The second term of Eq. (\ref{eps-G}) describes the spin flip excess
energy of 
forming $\sigma$ from $A+B$
already prepared at the volume $V$.
The correlation function
$\overline{\Pi}_f$ is defined 
as a product of the variables $\hat{S}_i$
over all sites of the figure $f$
with the overbar denoting an
average over the $D_f$ symmetry equivalent figures per lattice site.
Equation (\ref{eps-G}) is similar to
Eq. (\ref{ising}), but here the effective interaction energies
$\{\nu_f\}$ are {\em volume-independent} pure spin flip
energies.

We will examine the invertibility of the inverse approach
by performing a ``controlled experiment'':
As input, we use two ``exact'' sets of
interactions, $\{\nu_f\}$.
Equation (\ref{ising}) is then
used along with these $\{\nu_f\}$ 
in direct Monte Carlo (MC) simulations to obtain the ``exact''
quantities such as SRO parameters ${\alpha(n)}$, 
structural energy differences $\delta E(\sigma,\sigma')$
between configurations $\sigma$ and $\sigma'$, transition
temperaures $T_c$, and 
the mixing energy of the random alloy $\Delta H_{\rm mix}$.
We then contrast the results of this ``direct
procedure'' with those of the ``inverse procedure'', 
in which the
set $\{ \alpha(n) \}$ (obtained in the direct procedure
from the {\em known, exact} $\{\nu_f\}$) is used as input to
deduce the interactions $\{\tilde{\nu}_f\}$ by IMC simulations
from which we then obtain $\tilde{\alpha}(n)$, 
$\delta \tilde{E}(\sigma,\sigma')$,
$\tilde{T}_c$, and $\Delta \tilde{H}_{\rm mix}$.

We use as input two sets of interaction energies
$\{\nu_f\}$ (see Table \ref{comp.jf}).  
As an illustration of physically realistic interactions,
we use one set of interactions that was
recently extracted \cite{Wolverton95} 
from $T=0$ first-principles calculations of
formation energies of ordered fcc-based Ni$_{1-x}$V$_x$ compounds
and reproduces reasonably well many of the measured physical
properties.
This ``realistic set'', which we call Set 2, contains pair interactions
up to fourth neighbors, as well as three and four-body
interactions.
Set 1 is identical to 
Set 2, except that we have set equal to zero all multibody
interactions.

We first contrast the {\em directly calculated} alloy properties
using Sets 1 and 2 in MC simulations. 
For the direct MC calculations, a system size 
of 4096 atoms was used with periodic boundary 
conditions, 1200 Monte Carlo steps (MCS) were 
used for equilibration, and averages were 
typically taken over 1800 MCS.  Temperatures of 
$T$=850 K and 2300 K were used for the SRO 
calculations for Sets 1 and 2.  
Fig. \ref{alpha.k} shows the SRO
$\alpha({\bf k})$ calculated directly from 
$\{\nu_f\}$ for Sets 1 and 2 at composition $A_3B$.
Only multibody interactions
contribute to the difference
between the SRO of Sets 1 and 2, and this difference is 
dramatic:  $\alpha({\bf k})$
of Set 1 shows peaks at the X-points $\langle 100 \rangle$
whereas $\alpha({\bf k})$ of Set 2 shows peaks at the
W-points $\langle 1\frac{1}{2}0 \rangle$ as seen 
experimentally in Ni$_3$V. \cite{Caudron92}  
Table \ref{comp.jf} also shows that multibody
interactions change the 
ground state structure from $L1_2$ to the {\em observed}
\cite{Smith87} $D0_{22}$ structure and that
the energy difference between these two structures,
$\delta E(L1_2,D0_{22})$, 
changes from -4 to +103 meV/atom and $T_c$ changes from
630 to 1850 K upon inclusion of multibody interactions.
Also, note from Table \ref{comp.jf}
that  $G(x=1/4)$ is a significant fraction of 
the random alloy mixing enthalpy,
$\Delta H_{\rm mix}(x=1/4)$.
Thus, from the directly calculated values 
it is clear that both multibody interactions and
the elastic energy $G(x)$ are physically very important
in this alloy system.

Using the directly calculated $\alpha({\bf k})$, we now
apply IMC to recover the interactions energies.
Following the tradition among practitioners of the IMC method,
only pair interactions were retained
in the energy expression of IMC.  
First, configurations were produced which reproduced
the input values of 35 shells of $\alpha(n)$.  System sizes
of 262,144 and 216,000 sites were used for Sets 1 and 2.
IMC simulations were performed on three crystals compatible with the
sets of $\alpha(n)$, and averages were taken over these
three crystals.  Tests were performed of the convergence
of the inverse procedure with respect to the number of
pairs included:  Many different sets of
pair interactions were used (between 4 and 20 shells),
and from these calculations, the number of pairs needed to
adequately reproduce $\alpha(n)$ were determined.

%
%
\begin{figure}[tb]
\hbox to \hsize{\epsfxsize=1.00\hsize\hfil\epsfbox{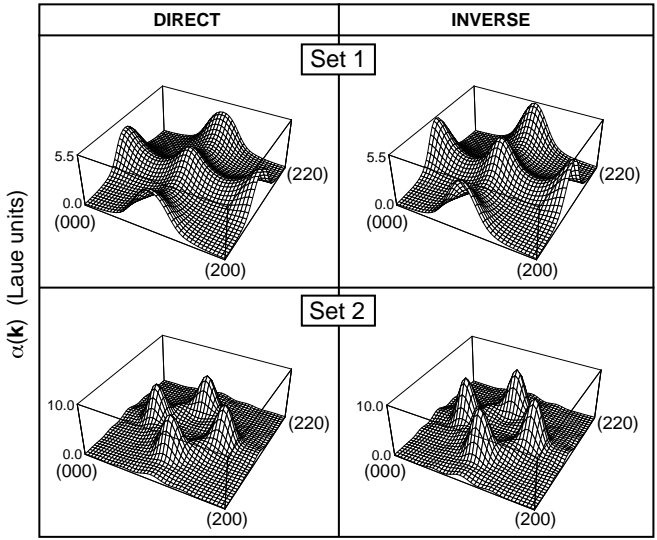}\hfil}
\nobreak\bigskip
\caption{Comparison of the (001) planes of
$\alpha({\bf k})$ and $\tilde{\alpha}({\bf k})$.}
\label{alpha.k}
\end{figure}

{\em Short-Range Order}:
Fig. \ref{alpha.k}
compares the recalculated $\tilde{\alpha}({\bf k})$ (computed from
$\tilde{\nu_f}$) and $\alpha({\bf k})$ (computed from $\nu_f$).
Both sets of SRO are well reproduced by the IMC procedure.
The accurate inversion of SRO 
[i.e., $\{ \tilde{\alpha}(n) \} \approx \{ \alpha(n) \}$]
has been demonstrated previously by many authors
(see, e.g., Refs. \onlinecite{Gerold87,Schweika88}).
However, in these previous studies, {\em measured} SRO was used as input
to the IMC, and thus the interactions which produced the input
SRO were not known.  We have shown that 
even when multibody interactions are used to produce
$\alpha(n)$ (as in Set 2), the IMC procedure (using pair
interactions only) reproduces this SRO quite well.

{\em Effective Interactions}:
We compare the values of $\nu_f$ vs. $\tilde{\nu}_f$ 
in Table \ref{comp.jf}.
For Set 1, the IMC algorithm 
closely reproduces the input set of pair interactions;
The standard deviation between $\{\nu_f\}$ and
$\{\tilde{\nu}_f\}$ for Set 1 is 2.7 meV/atom.
For Set 2, the direct and inverse sets of interactions
differ dramatically in several respects:  (i)  Three and four body
interactions are present in $\{\nu_f\}$, but are specifically 
excluded from the pair-only IMC calculation of $\{\tilde{\nu}_f\}$.
(ii)  There are huge differences in {\em pair} interactions (the standard
deviation of the first four pairs is 275 meV/atom).  
(iii)  Seven pair interactions
were required in the IMC to reproduce $\alpha(n)$,
whereas only four pair interactions were present in the direct set.
(A similar increase in range was reported in Ref. \onlinecite{Schweika89}.)
However, even though there are enormous differences between
$\nu_f$ and $\tilde{\nu}_f$ of Set 2, they both
produce nearly identical SRO patterns (see Fig. \ref{alpha.k}).
This surprising fact indicates
that even the {\em pair} interactions 
are not determined uniquely from a SRO pattern.
[This non-uniqueness was also found by Schweika and Carlsson 
(Ref. \onlinecite{Schweika89}; Fig. 3a), who
in contrast to the present work, used a high-temperature expansion 
(whereas we use IMC)
to invert SRO generated by pair and multibody interactions.]
We assert that due to the non-uniqueness of
pair interactions derived from IMC, they 
cannot generally be compared with other sets of pair interactions.
When multibody interactions are physically important,
the non-uniqueness of these sets make such comparisons meaningless.
For example, Schweika and Carlsson \cite{Schweika89}
found that inversion of SRO 
produced interactions
$\tilde{\nu}_f$ which were {\em temperature-dependent}
even though the input set $\{\nu_f\}$ was not.  Clearly,
this temperature-dependence is 
not due to {\em physical} effects (e.g., vibrational or
electronic excitation effects), but rather due
to the fact that a pair-only inverse scheme does not 
recover information on the multibody interactions
$\{\nu_f\}$.  

Structural energies, ground states,
and transition temperatures computed 
from $\{\nu_f\}$ are compared
with those computed from
$\{\tilde{\nu}_f\}$ in Table \ref{comp.jf}.

{\em Mixing Energies}:
Table \ref{comp.jf} also shows values of the mixing
enthalpy of the random alloy
$\Delta H_{\rm mix}(x=1/4)$.
For Set 1, where $\{\tilde{\nu}_f\} \approx \{\nu_f\}$, the
direct and inverse values of $\Delta H_{\rm mix}$ differ
by more than 100 meV/atom,
{\em even in cases (such as Set 1) dominated by pair interactions}.
In Set 2, the comparison of $\Delta H_{\rm mix}$ is even
worse (direct and inverse values differ by more than 350 meV/atom).
Deducing values of 
$\Delta H_{\rm mix}$ is clearly not reliable in the inverse
procedure.

We have seen that while that total energy $E(\sigma,V)$
{\em defines} the complete set of interaction energies,
inversion of quantities (e.g., SRO) other than $E(\sigma,V)$
may lead to a loss of information.
We now use Eq. (\ref{eps-G}) to distinguish different classes of alloy
properties and discuss which are invertible:

(a) Physical properties that depend on both $G(x)$ and on the spin
flip energies $\{\nu_f\}$ include any quantity which involves
the energetics of two or more concentrations and hence,
two or more volumes.  
[Note that $G(x)$ depends on $x$, but not on the particular
atomic arrangement (``configuration'') $\sigma$.]
Examples include
the formation energy of a structure [which involves $V$, $V_A$,
and $V_B$, c.f. Eq. (\ref{deltae})], 
the mixing energy $\Delta H_{\rm mix}$ of the random alloy,
and two-phase equilibria in a composition-temperature phase diagram.
Since ``type-(a)'' properties such as the set
$\{\Delta E(\sigma,V)\}$ contain complete information on both
$G(x)$ and on all $\{ \nu_f \}$,
given the measured or {\em ab-initio} calculated
energies $\{ \Delta E(\sigma,V) \}$,
it is possible to invert Eq. (\ref{eps-G}) and
in principle extract the ``exact'' 
$G(x)$ and $\{\nu_f\}$, as demonstrated in Ref. \onlinecite{Ferreira88}.  
Thus, inversion of ``type-(a)'' properties involves no loss
of information.

(b) Physical properties that do not depend on $G(x)$ include
energy differences
of iso-compositional configurations $\sigma$ and $\sigma'$,
$\delta E(\sigma,\sigma')$.
The order-disorder transition temperature $T_c$
at stoichiometric composition
also falls into this class since it
involves the energy difference between the disordered high-temperature
phase and the partially ordered low-temperature phase, both at the
same $x$. 
Another physical property which does not depend on $G(x)$
is the atomic SRO.
$\alpha(n)$ involves a competition between energies of
a random and a short-range ordered structure, both at
the same volume $V(x)$; therefore, $\alpha(n)$ (even
if determined for several compositions) contains
no information about $G(x)$.
Therefore, inversion of a ``type-(b)'' property, such as SRO,
cannot provide any information on $G(x)$, 
{\em even if the SRO covers a range of compositions}. Consequently, 
the interactions $\{\tilde{\nu}_f\}$ extracted from
such a procedure do not allow calculation of ``type-(a)'' 
properties, such as formation energies [Eq. (\ref{deltae})], mixing energies, 
or the phase-coexistence regions of the phase diagram.  
This point is highlighted by recent studies \cite{Lu94}
on Ni$_{1-x}$Au$_x$:  This is a phase-separating system;
however, the SRO is of ordering type.  Inverting the SRO will thus
inevitably produce ordering-type $\{\tilde{\nu}_f\}$,
which are useless for predicting the miscibility gap
phase diagram or the correct $\Delta H_{\rm mix} > 0$.
These conflicts are resolved \cite{Lu94} 
by using $G(x)$ in the Ising-like expansion.

We conclude that:  (i) The IMC algorithm provides a set
of {\em pair} interactions which accurately reproduces the input SRO 
whether or not {\em multibody} interactions are used to generate this
input.  When {\em only} pair interactions are involved, the
inverse procedure can even provide accurate values of 
structural energy differences, 
ordering energies, and energies of SRO; 
However, (ii) when multibody interactions are physically important, 
even the pair interactions are incorrectly determined by the
inversion of SRO.
The structural or ordering energies deduced
from the inverse procedure can thus contain substantial errors.  
(iii)  Finding a set of interactions which
reproduces a given set of SRO is found to be a non-unique process:
dramatically different sets of interactions (one set with
pairs only, one set with pairs and multibodies) may still
produce quantitatively the same SRO.  
Thus, comparing sets of interactions
from IMC with other sets of interactions may be unwarranted.
However, comparing a theoretical SRO pattern 
to a measured one is a sound procedure.
(iv) Inverting the SRO always removes information on energy
terms that are SRO-independent, e.g., $G(x)$.  This loss
prevents, in principle, the interactions deduced from SRO from
being applied to predict phase-coexistence regions of the phase
diagram or $\Delta H_{\rm mix}$.

This work was supported by the Office of Energy Research
(OER) [Division of Materials Science of the Office of Basic Energy
Sciences (BES)], U. S.  Department of Energy, under contract No.
DE-AC36-83CH10093.

\vskip -12 pt

\end{document}